# Through-thickness superconducting and normal-state transport properties revealed by thinning of thick film *ex situ* YBa$_2$Cu$_3$O$_{7-x}$ coated conductors

D.M. Feldmann, D.C. Larbalestier
*Applied Superconductivity Center, University of Wisconsin – Madison, Madison, Wisconsin 53706*

R. Feenstra, A. A. Gapud, J. D. Budai
*Oak Ridge National Laboratory, Oak Ridge, Tennessee 37831*

T.G. Holesinger, P.N. Arendt
*Los Alamos National Laboratory, Los Alamos, New Mexico 87545*

[Abstract]
A rapid decrease in the critical current density ($J_c$) of YBa$_2$Cu$_3$O$_{7-x}$ (YBCO) films with increasing film thickness has been observed for multiple YBCO growth processes. While such behavior is predicted from 2D collective pinning models under certain assumptions, empirical observations of the thickness dependence of $J_c$ are believed to be largely processing dependent at present. To investigate this behavior in *ex situ* YBCO films, 2.0 and 2.9 µm thick YBCO films on ion beam assisted deposition (IBAD) – yttria stabilized zirconia (YSZ) substrates were thinned and repeatedly measured for $\rho(T)$ and $J_c(H)$ . The 2.9 µm film exhibited a constant $J_c$(77K,SF) through thickness of ~1 MA/cm$^2$ while the 2.0 µm film exhibited an increase in $J_c$(77K,SF) as it was thinned. Neither film offered evidence of significant dead layers, suggesting that further increases in critical current can be obtained by growing thicker YBCO layers.

Coated Conductors (CCs), or second generation high temperature superconducting (HTS) wires, hold great promise for commercial applications and electric power applications, such as transmission lines, transformers and motors. To meet the technical requirements of these applications, CCs need to deliver large critical current ($I_c$) values.[1] An obvious approach to improving $I_c$ is to increase the thickness $t$ of the superconductor layer, but for several YBCO growth processes,[2-7] there is a strong decrease in the critical current density $J_c$ with increasing film thickness $t$.

Most knowledge of the thickness dependence of the critical current density of YBCO films comes from studies of multiple films produced with different YBCO layer thickness.[2,3,6,7] A complementary approach to investigate $J_c(t)$ was employed by Foltyn *et al.*[4] Using ion milling, thick YBCO films were successively thinned and $J_c$ measured at multiple thickness on a single sample. This second approach also produces a variable thickness dataset, however, with reduced emphasis on thickness dependent trends in the growth, but enhanced emphasis on through-thickness homogeneity. The YBCO in Ref. 4 was grown by pulsed laser deposition (PLD) on metal tapes buffered with yttria-stabilized zirconia (YSZ) textured by ion-beam-assisted deposition (IBAD). Strong structural degradation was observed in the films, most significantly in the form of "dead-layers" ($I_c=J_c=0$) at both the film surface and the buffer layer interface. Subsequent work by Foltyn and others has shown that these "dead" layers are not endemic of thick YBCO coated conductors, but instead dependent on the growth conditions[8-11]. In this letter, we report ion-milling studies on YBCO grown by the $BaF_2$ *ex situ* process on IBAD-YSZ buffered metal substrates.

The details of the BaF$_2$ *ex situ* growth process[7,12] and preparation of IBAD-YSZ buffered metal tapes[13] are given elsewhere. Briefly, precursor layers of the desired thickness were prepared by vacuum deposition using three electron-beam sources onto a "cold" substrate; the layers were subsequently converted into epitaxial YBCO by annealing in a furnace under flowing gas conditions at atmospheric pressure. Maximum temperatures during anneal reached 780°C and the growth rate was about 1-1.5 Å/s. The oxygen partial pressure was 200 mTorr and the water partial pressure (needed to decompose incorporated BaF$_2$ of the precursor layer) was kept below 5 Torr.

Two samples with YBCO layer thickness of 2.0 and 2.9 µm were prepared on substrates having similar but slightly different texture. The full-width-at-half-maximum (FWHM) values of the (205)/(025) YBCO reflections in x-ray diffraction (XRD) Φ-scans for the 2.0 and 2.9 µm films were 6.9° and 5.9° respectively. Bridges 300 µm wide were cut with a laser to restrict $I_c$ to < 10 A. The resistivity $\rho(T)$ and $J_c(H,77K)$ were measured using pulsed-current (50 ms pulse duration with 30ms voltage read) with a standard four-point configuration and a 1 µV/cm $J_c$ criterion. For ion milling, samples were cooled to ~230 K and Ar ions of energy 500eV impacted the samples at 45° with the film surface normal. The YBCO etch rate was ~12 nm/min. Thickness was measured using a Tencor profilometer.

Figure 1 shows $\rho(T)$ for the 2.9 µm film as it was thinned to a thickness *t*. The curves exhibit a slight positive curvature generally associated with O$_2$ over-doping[14]. The inset to Figure 1 shows $T_c$ as a function of thickness. The fluctuations in $T_c$ are believed to be greater than the measurement accuracy, but overall $T_c$ remains relatively constant. The resistivity $\rho(300K)$ as a function of film thickness (Figure 2) is nearly constant at ~400 µΩ-cm down to ~0.4 µm, below which it starts to increase abruptly, suggestive of a reaction between the YBCO and buffer layer. However, based on the constant $T_c$, $\rho(300K)$, and the curvature of the $\rho(T)$ data we conclude that repeated ion milling did not significantly damage the film or affect the oxygenation state of the remaining layer.

Also shown in Figure 2 is $I_c(77K)$ as a function of thickness, where the critical current of the bridge has been normalized to a 1-cm width (units: A/cm). The data exhibits a high degree of linearity, and provides no evidence of substantial dead layers in the film at either the substrate interface or the film surface.

Figure 3(a) shows $J_c(t,77K)$ for self-field (SF) and several applied fields ($H//c$). $J_c(SF)$ is nearly constant through thickness, tightly scattered around 1 MA/cm$^2$. Figure 3(b) shows data for the 2.0 µm thick YBCO film. Due to a slightly worse substrate texture, this film had a lower full-thickness $J_c(77K,SF)$ of 0.73 MA/cm$^2$. $J_c(t,SF)$ is not as flat as for the thicker-film sample with $J_c$ rising to almost twice the full-thickness value at 0.12 µm. Again, the $J_c(t)$ data of these samples give no evidence of dead layers or otherwise strong substrate reaction. This observation is consistent with the cross sectional scanning electron microscope (SEM) image of the 2.9 µm sample shown in Fig. 4. While there is more porosity in the upper portion of the film, the microstructure is dense through thickness with nothing to suggest a complete current blocking layer. An interesting question arising from Figs. 3(a) and 3(b) is why higher $J_c$ values were not recovered as the films were milled thinner. The solid line in the Figures is a power law fit of the form $J_c \propto 1/t^{1/2}$, fit to $J_c(t,SF)$ data from multiple films reacted from different precursor thickness.[7] At full thickness the $J_c(SF)$ of the 2.0 and 2.9 µm samples are representative of the fit, but as they are milled thinner $J_c(SF)$ deviates significantly. For instance, at a thickness of 0.32 µm, the 2.9 µm sample had a $J_c$ of 1.1 MA/cm$^2$ compared to 2.7 MA/cm$^2$ for a film reacted from a 0.35 µm precursor on a nominally identical substrate. For the *in-situ* PLD YBCO films of the study of Foltyn *et al.*,[4] the $I_c$ degradation in the bottom portion of the films was attributed to a reaction with the buffer layer that took place in the time it took to grow the upper portion of the films. Likewise, it is possible that a time-dependent reaction has degraded the bottom part of the *ex-situ* YBCO, although apparently to a lesser degree than in the PLD films. A second explanation, which is unique for the *ex-situ* process, is that the YBCO that nucleated at the substrate interface had poorer properties from the start (relative to films reacted from thinner precursors). For conversion of increasingly thicker films, the $H_2O$ required for the reaction and the HF produced must diffuse through increasingly more precursor material, and the thickness of the precursor itself becomes a growth parameter.

The flat $J_c(t)$ dependence observed for the 2.9 µm film would seem to be ideal and representative of a uniform microstructure. However, while the SEM shows a structure that is dense and relatively homogeneous through the thickness (Fig. 4) the TEM presents a different picture. A representative image is shown in Fig. 5. The bottom portion (~1.6 µm) of the film consists of large, well-formed YBCO grains (>10 µm) containing a high density of small, coherent $Y_2O_3$ particles. Some layers and discrete particles of Ba-Cu-O can also be found. The upper part of the film contains smaller, highly defective YBCO grains, and an overall different morphology and chemistry of second phases. The top part of the film appears similar to films reacted from thin precursors (< 0.5 µm). Thus, the image provides evidence of a "bi-layer" structure suggestive of different growth events through the thickness.

The absence of correlated behavior in the $I_c(t)$ (Fig. 2) and $J_c(t)$ (Fig. 3(a)) data for this film (near $t = 1.6$ µm) is highly surprising in view of this microstructure (Fig. 5). It suggests that either the different growth modes did not affect $J_c$ and the underlying fux pinning mechanism, or the defects that are controlling $J_c$ are not revealed by TEM at this magnification. The data clearly illustrates that it is possible to have a $J_c$ that is nearly independent of $t$, despite an inhomogeneous microstructure.

A further remarkable observation is that the empirical behavior $J_c \propto 1/t^{1/2}$ agrees with predictions from collective pinning models, based on a homogeneous random pinning potential.[15] If this fit were interpreted as intrinsic behavior then the $J_c(t)$ observed from thinning the 2.9 µm film should have followed more closely the $1/t^{1/2}$ fit. One explanation of this discrepancy, in the context of the collective pinning model, is that there is a reduced density of pins in the bottom part of the film. This interpretation is qualitatively consistent with the TEM image of Fig. 5 and would indicate that the primary difference between a high-Jc thin film and a thin remaining layer after ion milling results from a thickness dependent growth mechanism. Further work is in progress to test this hypothesis and to probe the dependence on growth conditions. We conclude that for a proper interpretation of thickness dependent effects, consideration should be given to the origin of flux pinning and the nature of defects in the superconductor.

In summary, we have directly observed of the *absence* of dead layers in thick (>2 µm) YBCO films grown on buffered metal tapes by successive ion milling and $J_c$ measurements. The lack of a surface dead layer is highly encouraging, and the linearity of the $I_c(t)$ curve of Figure 2 suggests that growing thicker *ex situ* YBCO films should lead to further increases in $I_c$. The roles of intrinsic and microstructural effects on $J_c(t)$ are not fully understood at this time.

Thanks to Darren Verebelyi (American Superconductor), Sang Il Kim (University of Wisconsin) and Alex Gurevich (University of Wisconsin) for experimental assistance and discussion. This work was supported by the AFOSR MURI, and benefited from the NSF funded MRSEC on nano-structured materials.

**Figure Captions**

FIG. 1. Temperature dependence of the resistivity for the 2.9 µm film at full thickness and selected values of $t$ after ion milling. Linear extrapolations to $T = 0$ K are shown for $t = 2.91$, 1.46, and 0.20 µm and have a slight negative intercept. Inset shows $T_c(t)$ for the film, where $T_c$ is defined as the highest temperature such that $\rho = 0$.

FIG. 2. Thickness dependence of $I_c$ (filled circles) and $\rho(300K)$ (open circles). The dashed line is a fit to the $I_c$ data, through the origin, with a slope of 103 A/(cm·µm).

FIG. 3. Thickness dependence of $J_c(77K)$ for (a) the 2.9 µm and (b) the 2.0 µm films. Solid lines are fits to $J_c$ data from multiple films reacted from different precursor thickness. With the exception of the $t = 0.19$ um data of the 2.0 µm film, the $J_c(t)$ curves maintain their qualitative shape in applied fields.

FIG. 4. Cross sectional SEM image of the 2.9 µm thick YBCO film.

FIG. 5. Cross sectional TEM image of the 2.9 µm sample, showing a distinct bi-layer structure.

**Figure 1**

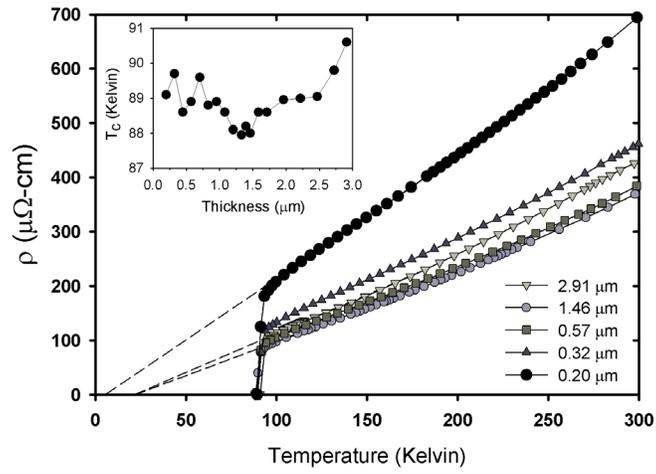

**Figure 2**

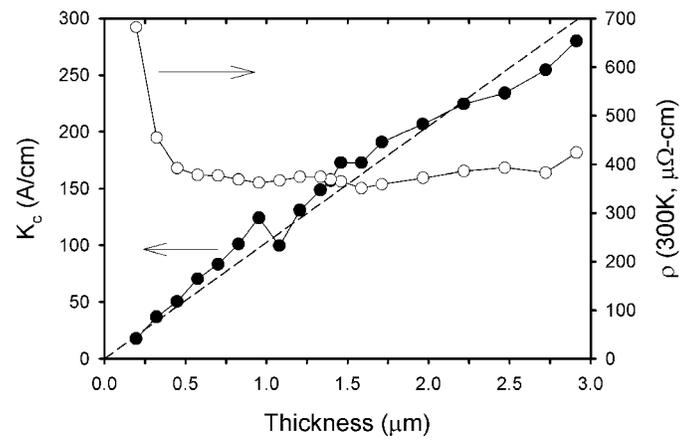

**Figure 3**

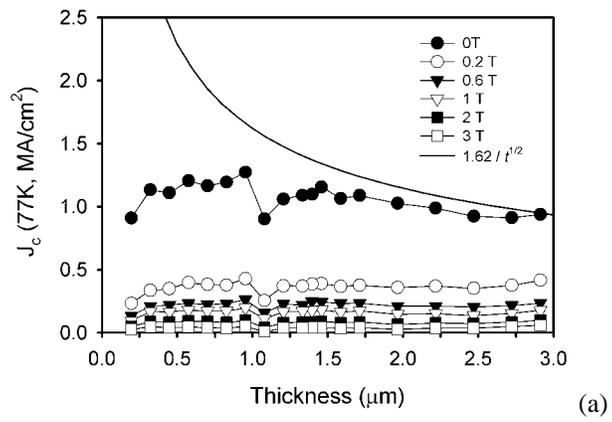

(a)

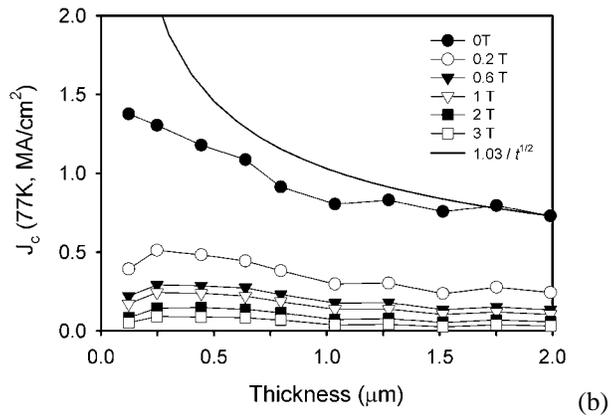

(b)

**Figure 4.**

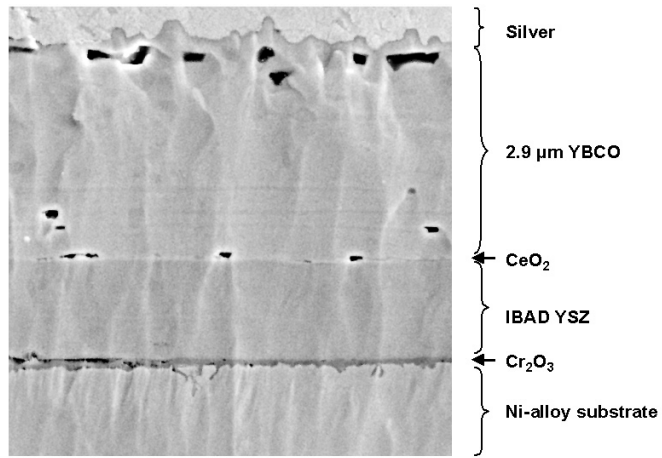

**Figure 5**

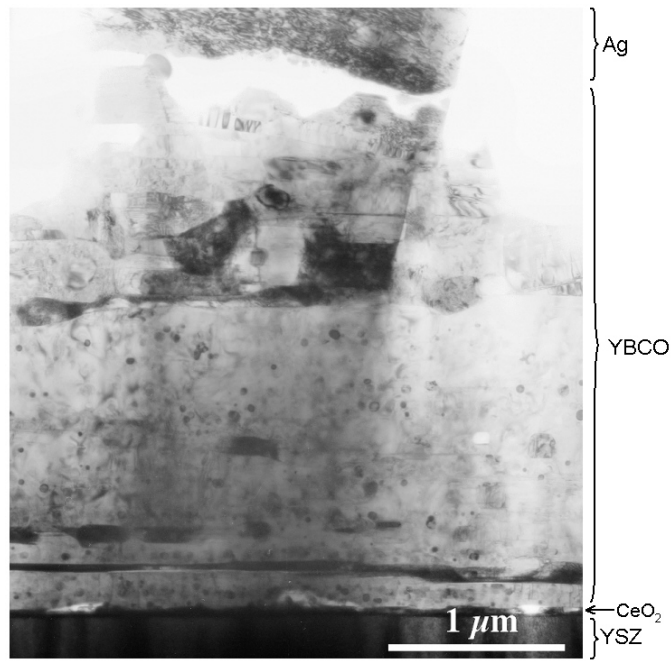